\documentclass[twoside]{ilcws08}
\usepackage[latin1]{inputenc}
\usepackage[dvips]{graphicx,epsfig,color}
\usepackage{wrapfig,rotating}
\usepackage{amssymb,amsmath,array}

\begin{document}
\title{
GARLIC - GAmma Reconstruciton for the LInear Collider} 
\author{Marcel Reinhard and Jean-Claude Brient
\vspace{.3cm}\\
LLR - Ecole polytechnique, IN2P3/CNRS \\
Palaiseau - France
}

\maketitle
\begin{abstract}
In order to profit from the high granularity of the calorimeters proposed for the ILC that are suitable for the Particle Flow Approach, specialised clustering algorithms have to be developped. GARLIC is such an algorithm with the goal to find and identify pointing photons in the electromagnetic calorimeter. This would help to improve the energy resolution on the photon contribution in jets.
\end{abstract}
\section{Introduction: The GARLIC algoithm}
The GARLIC algorithm is based on the REPLIC package. Its main idea is to make use of the typical shape of an electromagnetic shower. Due to its high density, a simple neighbour criterion is sufficient for clustering. In this way, noise hits as well as fluctuations in the extremities of the shower are suppressed and at the same time the two-particle separation is optimised. GARLIC has benn implemented in the MARLIN framework (\cite{MARLIN}) and is adapted to the Si/W ECAL, both for the CALICE prototype as well as for the full detector model (LDC/ILD). To account for the differences in geometry, two versions of the algorithm exist. They are identical in terms of the actual clustering but adapted to the respective geometry. The main steps in the algorithm are:
\begin{itemize}
 \item Recursive pre-clustering based on the distance between hits
 \item Seed search via two-dimensional projection of the energy deposit in the first $7X_0$
 \item Core building along direction found by the seed search
 \item Iterative clustering based on neighbour criterion, proceeding from front to back
 \item Rejection via simple criteria (number of hits, minimum energy, seed criteria)
 \item Correction for energy losses in wafer guard rings and inter-module gaps
 \item Rejection with neural network output based on cluster variables such as eccentricity, width, direction, energy deposit in different parts of the cluster, etc.
\end{itemize}

\section{Characterisation of the prototype version}
The CALICE Si/W ECAL prototype has been tested in several test beams. GARLIC can be applied on data taken with electron beams to validate the clustering of electromagnetic showers. The basic concept of the algorithm works well. At a beam energy of 10 GeV at normal incidence, the mean fraction of clustered hits per event is about $97\%$, while the mean fraction of clustered energy is superior to $99\%$. This means that mainly low energy hits in the outer parts of the shower which represent fluctuations or noise hits are removed from the event. 
\subsection{Event cleaning}
Another important aspect of GARLIC is its capability to clean event samples in the prototype test beam. To reject pion background a simple variable can be defined that measures the ratio of the energy deposit in the core of a shower to the total energy of the cluster. The core is hereby defined as all the hits with a distance of less than $\sqrt{2} \times cellsize$ to the main axis of the cluster. This method proves as efficient as the rejection with a \v{C}erenkov counter. Since this detector was not available during all the data taking periods, GARLIC can be used to clean the event samples in these cases. Combining the two methods gives even better results. A second application for event cleaning is to require that exactly one cluster has been found in the event. This will reject multi-particle events as well as those where the incoming particle has started its shower upstream of the ECAL.
\subsection{Gap correction}
The Si/W ECAL shows some inefficient areas due to the presence of guard rings around the silicon wafers and the carbon fiber structure between modules. Detailed information can be found in \cite{AMpaper} and \cite{response_paper}. GARLIC's in-built gap correction has been kept very simple in order to keep it independent of the angle of incidence. So-called ghost hits are introduced in a gap between two hits. An energy deposit equal to the mean of the adjacent real hits, scaled by the fraction of the passive area to the area of one cell is attributed to this ghost hit. By applying this method losses can be reduced by about $5\%$ for events in the inter-wafer gaps and by about $8\%$ at the crossing of two such gaps. 
\subsection{Impact on linearity and resolution}
It turns out that the clustering has only small effects on the linearity and resolution of the detector. As mentioned above more than $99\%$ of the recorded energy is clustered. The main impact is then due to the gap correction that artificially increases the detector response at higher energies. To be consistent with the LDC/ILD version where the conversion from MIP to GeV has to be valid over a very big energy range and which induces a small non-linearity, the response is fitted with a parabolic function. As expected, the second order term turns out to be very small. The impact on the energy resolution is also small. The stochastic term \textit{s} in the parametrisation $\frac{\Delta E_{meas}}{E_{meas}} = \frac{s}{\sqrt{E(GeV)}} \oplus c$ changes by about $0.1\%$ (absolute) and the two values agree perfectly within the errors.
\section{Characterisation of the LDC/ILD version}
The LDC/ILD version has some more feautures than the prototype version, since it has to be adapted to a complete detector geometry and on more complex event types.
\subsection{Rejection of clusters from charged hadrons}
Separating clusters of electromagnetic showers from those created artificially, especially by charged hadrons that started interacting in the ECAL, needs several steps. The first and simplest one is to reduce the number of seeds that could be found from interacting charged hadrons. Therefore, tracks from charged particles are extrapolated into the ECAL volume and hits close by are removed from the pool of hits used for seed finding. Since only hits in the first $7X_0$ are taken into account only early showering hadrons will generate seeds. With hadrons at 10 GeV, in only $13\%$ of the events a seed is found. To remove the residual artificial clusters, a neural network has been trained with several variables describing the cluster properties.
\begin{figure}[h]
\centering
\includegraphics[scale=.45]{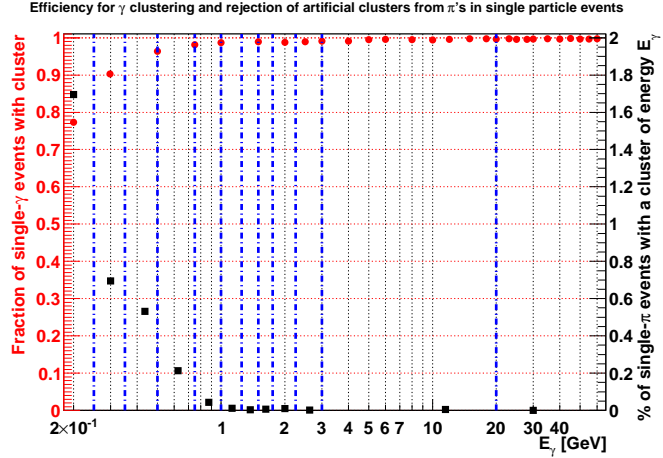}
\caption{Clustering efficiency for single $\gamma$-events (red dots) and percentage of events with artificial clusters from showers of single-$\pi$'s (black squares) in function of the energy of the simulated $\gamma$ and energy of the artificial cluster respectively}
\label{Fig:Efficiencies}
\end{figure}
\subsection{Clustering efficiency and purity}
Simulated single-$\gamma$ (0.2-150 GeV) and single-$\pi$ (1-110 GeV) events have been used to study GARLIC's clustering efficiency and the redidual fake clusters from hadron showers. In both cases the particle is required to enter the ECAL volume and not to interact in the tracker region. Figure~\ref{Fig:Efficiencies} summarizes these results. The dashed blue vertical lines indicate the cluster energies that are grouped for the neural network decision. The red dots give the fraction of events with at least one cluster found, as a function of the the $\gamma$-energy in single-photon events. The lower energy limit for accepted clusters is 150 MeV. At 500 MeV the efficiency to find a cluster is already $96\%$, at 1 GeV it is $98.6\%$ and larger than $99\%$ for energies bigger than 1.5 GeV. The black squares give the percentage of single-$\pi$ events that show a fake cluster of energy $E_{\gamma}$, while averaging over all simulated $\pi$ energies. One can see that it is very unlikely to create fake clusters above 1 GeV. At lower energies however, there are only few hits available to characterize the cluster, so selecting a fake cluster is more likely. A very important role is played by interactions in the tracker region. Rejection of the resulting fragments, that do not have any tracking information, proves near impossible. When allowing such interaction in the tracker region the background level more than doubles.
\begin{figure}[h]
\begin{minipage}[c]{0.55\columnwidth}
\centering
\includegraphics[scale=.37]{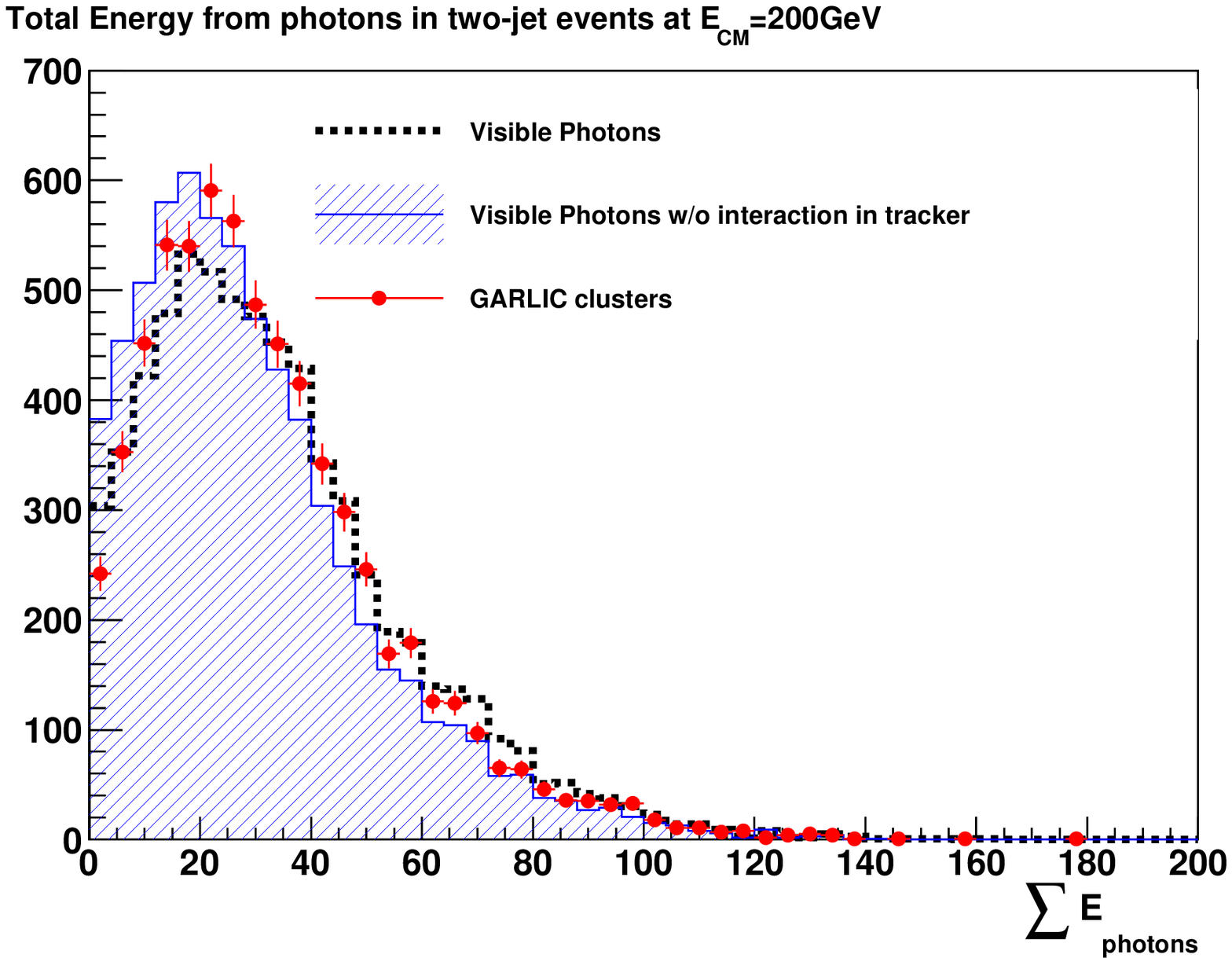}
\caption{Total energy deposit from photons in two-jet events at $E_{CM} = 200 GeV$. The definition of visible photons can be found in the text.}
\label{Fig:gamma_energy}
\end{minipage}
\hspace{3mm}
\begin{minipage}[c]{0.35\columnwidth}
\centering
\includegraphics[scale=.25]{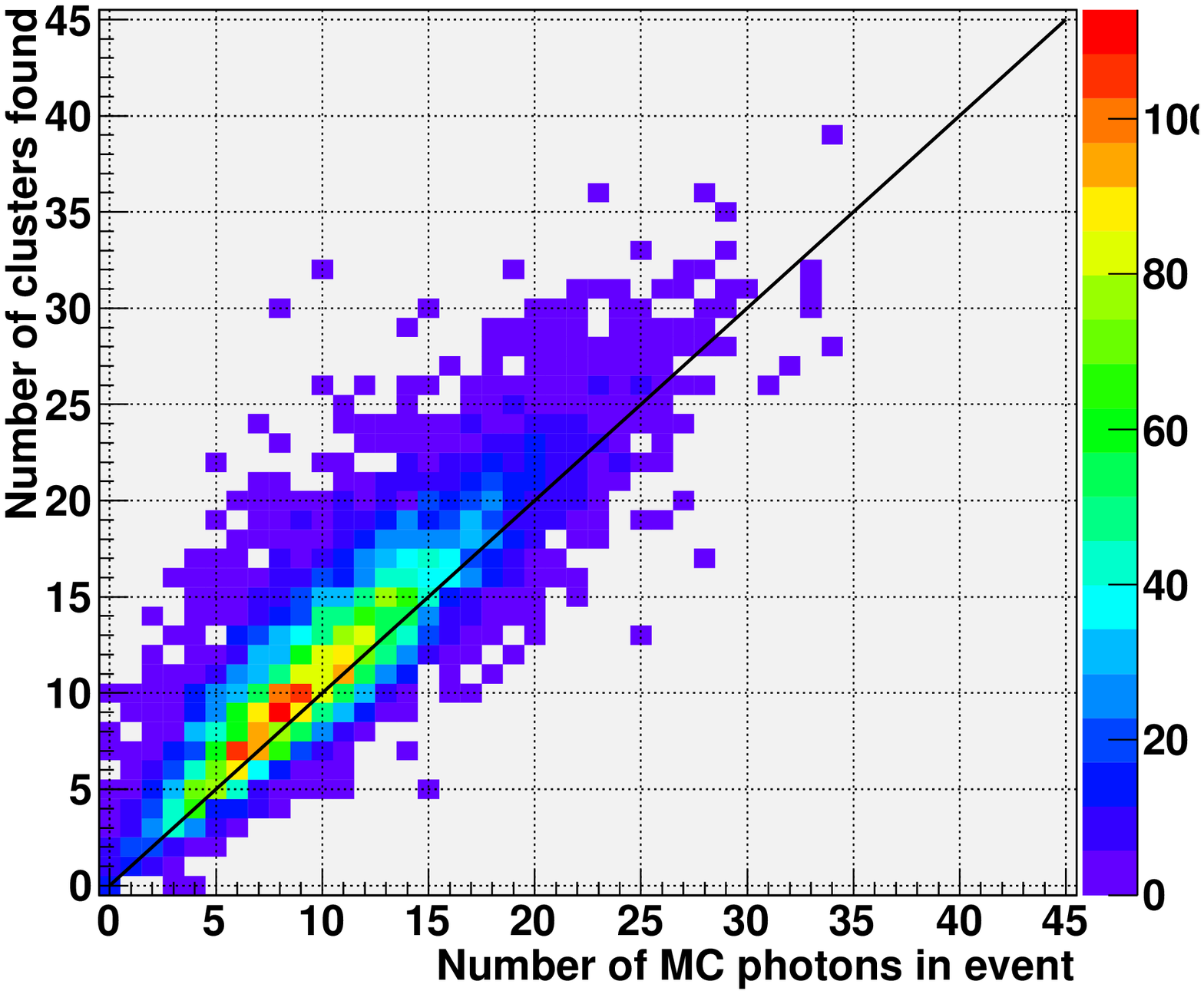}
\caption{Correlation between number of clusters found by GARLIC and simulated photons in two-jet events at $E_{CM} = 200 GeV$}
\label{Fig:photon_corr}
\end{minipage}
\end{figure}
\subsection{Application on jets}
About $30\%$ of the energy in jets is coming from photons. One of the goals of GARLIC is to optimise the resolution of this part of the jet energy and to isolate the photon clusters so following clustering algorithms can complete the particle flow by treating the charged particles and neutral hadrons. Two-jet events in the ILD detector at $E_{CM} = 200 GeV$ have been simulated and reconstructed with GARLIC. Figure~\ref{Fig:gamma_energy} shows the sum of the energy deposited by all photons in the two jets. Photons are defined as visible (for GARLIC) if they have a minimum energy ($E_{\gamma}>150 MeV$) and if they enter the ECAL volume ($|(cos\theta)|<0.9774$). One can see that the photon energy is quite well reconstructed. The correlation between the number of photon clusters and simulated photons as specified above is given in Figure~\ref{Fig:photon_corr}. As one can see the number of photons in the jet is often overestimated. The combination of these two plots suggest that there is still a significant number of artificial cluster background at very low energies. This is amplified by the creation of fake clusters due to hadron interaction as described in the treatment of single-particle events.
\section{Conclusions}
The study on test beam data shows that GARLIC does not have any negative influence on the energy resolution. The built-in gap correction helps to reduce the effect of energy losses in non-sensitive areas of the calorimeter. To optimise this correction different approaches are being studied. GARLIC works for photons down to 150 MeV with high efficiencies and low background. Nevertheless hadron interaction in the tracker region degrades its (and the particle flow) performance. 


\begin{footnotesize}




\begin{thebibliography}{99}
\bibitem{url} Presentation: \\ 
\verb$http://ilcagenda.linearcollider.org/contributionDisplay.py?contribId=135&sessionId=23&confId=2628$

\bibitem{MARLIN} \verb$http://ilcsoft.desy.de/portal/software_packages/marlin/index_eng.html$
\bibitem{AMpaper} J.~R\'{e}pond {\it et~al.}, JINST:3 P08001 (2008).
\bibitem{response_paper} The CALICE Collaboration, arXiv:0811.2354v1 (2008).

\end{thebibliography}
%

\end{footnotesize}


\end{document}